\documentclass[doublecol]{epl2}
\usepackage{graphicx}
\usepackage{latexsym}
\usepackage{amsmath}
\usepackage{amssymb}
\usepackage{epstopdf}
\usepackage{enumerate}
\usepackage{dcolumn}
\usepackage{bm}

\begin{document}
\title{ Exciton
condensation
in thin film topological insulator}

\author{Eun Gook Moon\inst{1,2}, Cenke Xu\inst{2}}

\institute{
  \inst{1} Department of Physics, Harvard University, Cambridge
MA 02138, USA\\
  \inst{2} Department of Physics, University of California, Santa Barbara, California 93106, USA
}

\pacs{nn.mm.xx}{First pacs description}
\pacs{nn.mm.xx}{Second pacs description}
\pacs{nn.mm.xx}{Third pacs description}

\abstract{
We study the many-body physics in thin film topological band
insulator, where the inter-edge Coulomb interaction can lead to an
exciton condensation transition. We investigate the universality
class of the exciton condensation quantum critical point. With
different chemical potentials and interactions, the exciton
condensation can belong to $z = 2$ mean field, or 3d XY, or
Yukawa-Higgs universality classes.
%
%
The interplay between exciton condensate and the time-reversal
symmetry breaking is also discussed.
Predictions of our work can be tested experimentally by
tuning the chemical potentials on both surfaces of the thin film
through gate voltage.
We also show that
all the analysis of the exciton condensate can be directly applied
to a spin-triplet superconductor phase with attractive inter-edge
interaction.}

\date{\today}

\maketitle


Quantum phases protected by topology have shown enormous
interesting behaviors.
Despite of unusual quantized responses to
external fields \cite{qi2008,andrew2008}, topological phases
usually manifest themselves with their stable edge states. For
instance, the three dimensional topological band insulator (TBI)
is characterized by its single Dirac cone edge state, with the
Dirac point located at the time-reversal ($\mathcal{T}$) invariant
point in the edge Brillouin zone. These edge states were predicted
theoretically, and also observed very successfully experimentally
in materials such as $\mathrm{BiSb}$, $\mathrm{Bi_2Se_3}$ and
$\mathrm{Bi_2Te_3}$
\cite{hasan2009a,hasan2009b,fang2008,zxshen2009,fu2007,fu2008}. It
was understood that $\mathcal{T}$ is crucial to the stability of
the edge states \cite{kane2005a,kane2005b,fu2007,fu2007a},
basically because one cannot open up a $\mathcal{T}-$invariant
Dirac mass gap for a single edge. Enormous interests were devoted
to the $\mathcal{T}$ breaking at the edge states of TBI, including
the cases with $\mathcal{T}$ broken by magnetic impurities and
broken spontaneously due to interactions
\cite{xuedge,wuedge,xuzhang,xuhelical1,xuhelical2,yaoran2010}.

In a thin film sample of TBI, since the two edges are close enough
to interact with each other, $\mathcal{T}$ is no longer sufficient
to protect the stability of the edge states $i.e.$ it is allowed
to open up a $\mathcal{T}$ invariant gap for both edge states.
Recently, the gap at the edge states was indeed observed in
experiments on thin film TBI, when the thickness of the film is
small \cite{xue1,xue2,cho}.  
It was proposed that even without direct
inter-edge tunnelling, the local Coulomb interaction can also gap
out the edge states through exciton condensation
\cite{SMF} $i.e.$ a particle-hole pair bound state across the thin
film condenses. This effect is most prominent when a specific
biased gate voltage is applied to two edges, where there is a
``nesting" between two Fermi surfaces, and the ``exciton"
susceptibility diverges~\cite{SMF}. Lately it was proposed that
large dielectric constants of these materials increases the layer
separation range over which the inter surface coherence
survives~\cite{Hankiewicz}. In our current paper we hope to go
beyond the mean-field consideration in Ref.~\cite{SMF} and study
the critical properties of the exciton condensate physics.

\begin{figure}[tb]
\begin{center}
    \includegraphics[height=.25\textwidth]{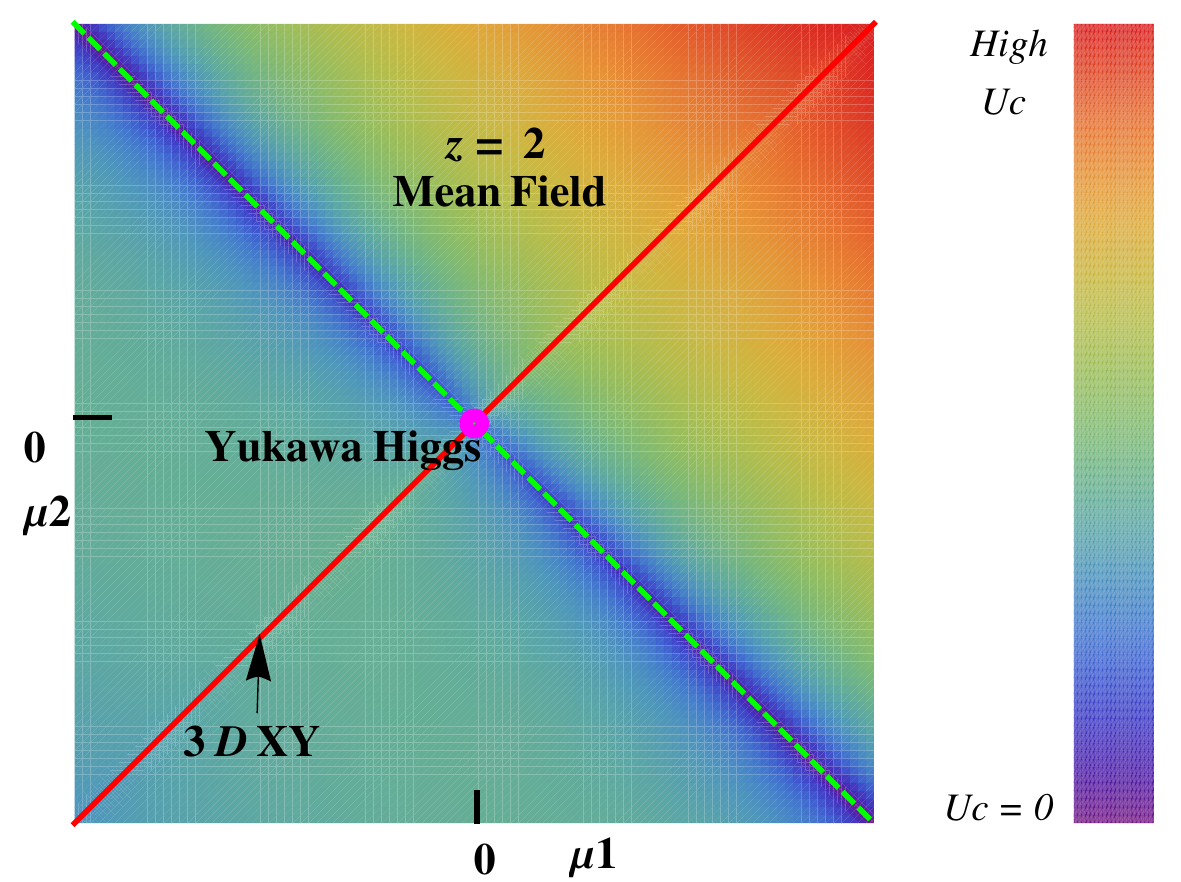}
 \caption{The phase diagram of Eq.~\ref{hamiltonian}. The
dark/bright colors denote the low/high critical Coulomb
interaction.} \label{phasediagram}
\end{center}
\end{figure}

We illustrate our main results of this paper in the phase diagram
(Fig.~\ref{phasediagram}) plotted against the chemical potentials
of the two edges, which we assume can be separately tuned with
gate voltages on both edges. The color in this phase diagram
denotes the critical Coulomb interaction $U_c$ required to drive
the exciton condensate, and at the line $\mu_1 + \mu_2 = 0$ except
for the origin, $U_c = 0$ due to the divergence of exciton
susceptibility. In most area of this phase diagram, the quantum
critical point (QCP) belongs to the $z = 2$ mean field
universality class. At the origin $\mu_1 = \mu_2 = 0$, the
transition is described by the Yukawa-Higgs theory; and at the
special line $\mu_1 = \mu_2 \neq 0$, the transition belongs to the
3d XY universality class. The same phase diagram and universality
classes apply to the superconducting transition with attractive
inter-edge ``Coulomb" interaction.

The effective Hamiltonian of the surface states reads \begin{eqnarray}
\mathcal{H}= \sum_{l=1,2} \psi^{\dagger}_l (v_l  \vec{\sigma}
\cdot \vec{p} - \mu_{l}) \psi_l + U n_1 n_2, \label{hamiltonian}
\end{eqnarray} where $v_l = (-1)^{l+1} v_f$ because the two edges of the
TBI have opposite helicities. This Hamiltonian clearly has
time-reversal symmetry $\mathcal{T}: \psi_l \rightarrow
i\sigma^y\psi_l$, $\vec{k} \rightarrow -\vec{k}$. There is also a
inversion symmetry $\mathcal{I}: \psi_1 \leftrightarrow \psi_2$,
$\vec{k} \rightarrow -\vec{k}$ when $\mu_1 = \mu_2$, and the
inversion is also a generic symmetry of materials such as
$\mathrm{Bi_2Te_3}$ and $\mathrm{Bi_2Se_3}$. Gate voltages
determine the relative chemical potentials, $\mu_{l}$. The second
term of Eq.~\ref{hamiltonian} describes the short ranged
inter-edge Coulomb interaction. We assume the screening length of
Coulomb interaction is always larger than the thickness of the
think film, therefore the inter-edge Coulomb interaction is still
important even when the direct inter-edge electron tunnelling is
ignorable. The intra-edge Coulomb interaction is also tentatively
ignored in this Hamiltonian, its effects will be discussed later.

Without inter-edge tunnelling, electron number in each layer is
independently conserved. This enlarged $\mathrm{U(1)\times U(1)}$
symmetry is spontaneously broken down to the diagonal U(1)
symmetry by exciton condensation with order parameter $\phi \sim U
\langle \psi^{\dagger}_{1} \psi_{2} \rangle$, which also lowers
the energy of the system. Then a mean field Hamiltonian can be
obtained from the Hubbard-Stratonovich transformation of the
original Hamiltonian Eq.~\ref{hamiltonian} \cite{SMF}: \begin{eqnarray}
H_{MF} = \sum_{l=1,2} \psi^{\dagger}_l (v_l  \vec{\sigma} \cdot
\vec{p} - \mu_{l}) \psi_l \nonumber + ( \phi^{*} \psi^{\dagger}_1
\psi_{2} + H.c.) +\frac{ |\phi|^2}{U} . \label{MF}\end{eqnarray} The
complex order parameter $\phi$ is invariant under $\mathcal{T}$,
but becomes its complex conjugate under inversion $\mathcal{I}$.
Clearly there are also other mean field channels of the Coulomb
interaction such as $\langle \psi^\dagger_1 \sigma^z \psi_2
\rangle$, but the exciton order $U \langle \psi^{\dagger}_{1}
\psi_{2} \rangle$ has the lowest mean field energy because it
opens up a Dirac mass gap, therefore we will focus on this order
parameter in our current work.

\begin{figure}[tb]
\begin{center}
    \includegraphics[width=3.2 in]{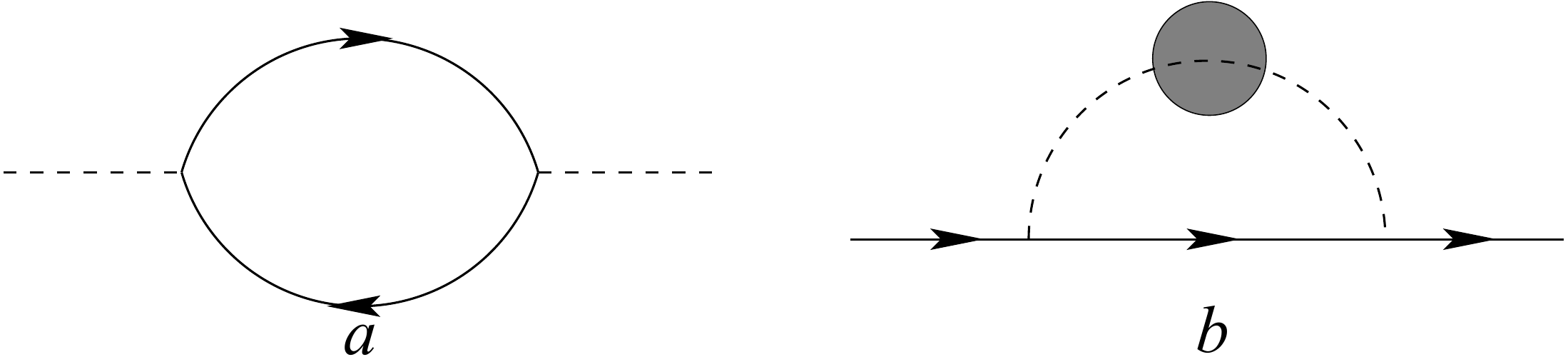}
\caption{$a$, one-loop Feynman diagram for renormalized Boson
Lagrangian, $b$, Feynman diagram for the fermion self-energy. The
dashed lines can denote either the exciton $\phi$ or exciton
bilinear $|\phi|^2$. The solid line denotes the fermion
propagator.} \label{bosonself}
\end{center}
\end{figure}

After integrating out the fermions, the static and uniform
renormalization to the mean field Hamiltonian of $\phi$ is
$\mathcal{L}_{eff} = (\frac{1}{U} - \chi_0)|\phi|^2$, $\chi_0$ is
the susceptibility of order parameter $\phi$ which can be
evaluated from Boson self-energy loop diagram
Fig.~\ref{bosonself}: \begin{eqnarray} \chi_0  = \Sigma_{\phi} (0,0) =
-\int_{i\omega,k} \mathrm{Tr}(\hat G_1(k,i\omega) \hat
G_2(k,i\omega)), \end{eqnarray} where $\hat{G}_l$ is Green's function of
fermion. This integral increases linearly with the ultraviolet
cut-off $\Lambda$ $i.e.$ it is not just a Fermi surface effect.
For instance, when $\mu_1 = \mu_2 = \mu$, $\chi_0 \sim \Lambda -
|\mu|$. Had we included the other mean field order parameter
$\psi^\dagger_1 \sigma^z \psi_2$, since the susceptibility of this
order parameter only comes from the Fermi surface, it would never
beat the order parameter $\phi \sim U \langle \psi^{\dagger}_{1}
\psi_{2} \rangle$ under consideration, as long as the size of the
Fermi surface is small compared with the UV cut-off.

\begin{figure}[tb]
\begin{center}
     \includegraphics[height=.27 \textwidth]{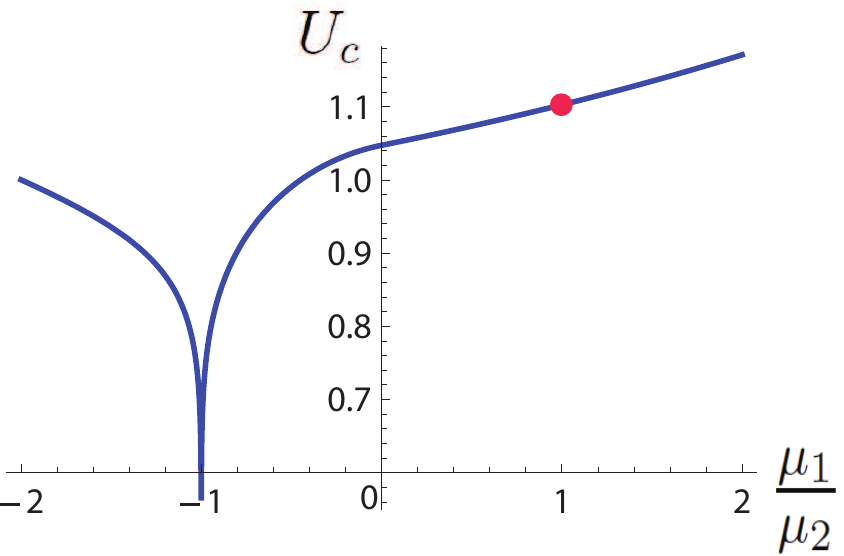}
\caption{The plot of critical Coulomb interaction $U_c$ against
chemical potential $\mu_1$, with fixed $\mu_2 > 0$. Critical
Coulomb interaction is measured in units of the $U_c$ with
$\mu_1/\mu_2 = -2$. $U_c$ is zero at $\mu_1 = - \mu_2$ due to the
logarithmically divergent exciton susceptibility. Everywhere but
$\mu_1 = \pm \mu_2$, the QCP have $z = 2$ dynamical exponents,
with mean field universality class. The special point $\mu_1 =
\mu_2 \neq 0$ (solid circle) has $z = 1$ dynamical scaling, and it
belongs to the 3d XY universality class.} \label{critical}
\end{center}
\end{figure}

If we fix $\mu_2$, the critical Coulomb interaction $U_c$ is
plotted in Fig.~\ref{critical}. As we mentioned, $U_c$ itself is
cut-off dependent. However, the relative difference between $U_c$
is UV cut-off independent, hence we can still compare $U_c$ with no ambiguity.
As we can see, at $\mu_1 = - \mu_2 \neq 0$, the critical Coulomb
interaction is zero, due to the fact that the exciton
susceptibility diverges logarithmically at this point. The
logarithmic divergence is a consequence of the Fermi surface
nesting between the two edges \cite{SMF}. At $\mu_1 = \mu_2 \neq
0$, although the two Fermi surfaces still have the same size, the
wavefunction overlap $\langle \psi_{1,\vec{k}}|\psi_{2,\vec{k}}
\rangle = 0$ for any momentum $\vec{k}$ at the Fermi surface,
therefore the susceptibility is not divergent. This matrix element
suppression is due to the opposite helicity of the two edges, and
as we will see, this suppression will also affect the dynamics and
universality class of QCP.

Now let us go beyond the mean field formalism and move on to the
universality class of the QCPs. Without fermions, the exciton
condensation would certainly belong to the 3d XY universality
class, while coupling to Fermions will likely modify the
universality class. Let us first discuss the case with $\mu_1 =
\mu_2 = 0$. At this point, after redefining $ \psi_2 \rightarrow
\sigma_z \psi_2$, the phase transition is described by the
following Lagrangian:
\begin{eqnarray} \mathcal{L}_{Higgs} &=& \bar{\psi}_l
\gamma^{\mu}
\partial_{\mu} \psi_l
+ \lambda \vec{\phi}\cdot \bar{\psi}\vec{\eta}\psi \nonumber
\cr\cr &+& |\partial_{\tau} {\phi}|^2 +v_{\phi}^2 |\nabla
{\phi}|^2 + r |{\phi}|^2 + \frac{u}{4} |{\phi}|^4 + \cdots \cr\cr
&& \vec{\eta} = (\eta^x, \ \eta^y), \  \vec{\phi} =
(\mathrm{Re}[\phi], \ \mathrm{Im}[\phi]). \end{eqnarray} Here $\ \gamma_\mu
= (\sigma^z, \sigma^x, \sigma^y)$,
$\eta^x$ and $\eta^y$ are two Pauli matrices that mix the two edges.
This model becomes precisely
the Higgs-Yukawa model which describes the Chiral symmetry
breaking of Dirac fermion, at least when $v_f = v_\phi$.
The transition of $\vec{\phi}$ is not 3d XY transition because the
coupling $\lambda$ is relevant at the 3d XY fixed point, based on
the well-known scaling dimensions $[\psi] = 1$, and $[\vec{\phi}]
= (d - 2)/2 + \eta/2 = 0.519$ at the 3d XY fixed point
\cite{vicari2003}. The critical exponents of this transition with
large $N$ have been calculated by means of $1/N$ and $\epsilon = 4
- d$ expansions
\cite{zustin1991,petersson1994,gracey1991,gracey1992}, and a
second order transition with non-Wilson-Fisher universality class
was found. Therefore we conclude that the transition is still
second order, with different universality class from the 3d XY
transition.

If $\mu_1 \neq \mu_2$, due to the mismatch of the size of the
Fermi surface, the loop diagram Fig.~\ref{bosonself} will not lead
to singular behavior for Boson self-energy at low frequency and
small momentum. However, the fermi surface mismatch breaks the
inversion symmetry of the Hamiltonian Eq.~\ref{hamiltonian},
therefore the following term is allowed in the Lagrangian: \begin{eqnarray}
\mathcal{L}_1  = h (i\phi^x \partial_\tau \phi^y - i\phi^y
\partial_\tau \phi^x)  \sim h \phi^\ast \partial_\tau \phi, \ h \sim \mu_2 - \mu_1. \label{z2}\end{eqnarray}
$\mathcal{L}_1$ leads to a $z = 2$ dynamical exponent, which is
analogous to the Mott insulator (MI) -superfluid transition in
Bose Hubbard model away from the tip of MI lobe
\cite{bosehubbard}, and also the XY magnetic transition in
magnetic field. In two spatial dimension, the $z = 2$ transition
is a mean field transition with marginally irrelevant
perturbations.

If $\mu_1 = \mu_2 \neq 0$, then the exciton condensation is
similar to a ferromagnetic transition in Fermi liquid, which
usually has $z = 3$ over damped quantum critical mode
\cite{hertz1976,millis1993}. However, in our case there is the
matrix element suppression effect mentioned before, namely: \begin{eqnarray}
|\mathcal{M}_{k,k+q}|^2 &=& |\langle k+q ,1 |
\psi^{\dagger}_{1,k+q} \psi_{2,k} |k,2 \rangle |^2 \cr\cr &=&
\sin^2 (\theta_{k+q} -\theta_k) \sim q^2/k_f^2. \end{eqnarray} This matrix
element suppression strongly affects the low energy dynamics of
the quantum critical fluctuations. For instance, the damping rate
of the quantum critical modes due to particle-hole excitations can
be calculated through the Boson self-energy diagram
Fig.~\ref{bosonself}$a$: \begin{eqnarray} &&
\mathrm{Im}[\Sigma_{\phi}(\omega, q)] \sim g^2 \int
\frac{d^2k}{(2\pi)^2} [f(\epsilon_{k+q}) - f(\epsilon_{k})]\cr\cr
&& \times \delta(|\omega| - \epsilon_{k+q} + \epsilon_k)
|\mathcal{M}_{k,k+q}|^2 \sim g^2 \frac{|\omega| q}{ v_fk_f^2}.
\label{damp}\end{eqnarray} This term will not lead to over-damped $z = 3$
quantum critical modes. The same effect was noticed in
Ref.~\cite{xuhelical1} in the context of $\mathcal{T}$ breaking at
the edge state of TBI, and following the argument of
Ref.~\cite{xuhelical1}, we can conclude that this transition still
belongs to the 3d XY universality class even though the order
parameter $\vec{\phi}$ couples linearly to Fermi surface in
Eq.~\ref{MF}.

The quantum critical modes also modify the Fermion's self-energy.
Using the Feynman diagram Fig.~\ref{bosonself}$b$, the Fermion
self-energy reads \begin{eqnarray} \Sigma(i \omega) &\sim& \int d^2k
d\epsilon G(i\epsilon + i\omega, \vec{k}) \langle
\vec{\phi}_{i\epsilon, \vec{k}} \vec{\phi}_{-i \epsilon, -\vec{k}}
\rangle |\mathcal{M}_{0,k}|^2. \label{psiself1}\end{eqnarray} Here we use
the full correlation at the 3d XY fixed point: $ \langle
\vec{\phi}_{i\epsilon,\vec{k}} \vec{\phi}_{-i\epsilon, -\vec{k}}
\rangle \sim (\epsilon^2 + v_\phi^2 k^2)^{-1 + \eta/2}$, $\eta$ is
the anomalous dimension of the order parameter $\vec{\phi}$ at the
3d XY fixed point. The leading order contribution to the imaginary
part of Fermion self-energy reads \begin{eqnarray}
\Sigma(\omega)^{\prime\prime} \sim (\lambda)^2 |\omega|^{2 +
\eta}\mathrm{sgn}[\omega] \ll |\omega|. \end{eqnarray} Therefore the
Fermion self-energy correction is always dominated by the linear
frequency term of the free electron propagator, the linear
coupling $\lambda$ does not destroy the Landau quasiparticles.

In addition to the linear coupling Eq.~\ref{MF}, another quadratic
interaction is also allowed by symmetry, but was omitted in the
mean field Hamiltonian Eq.~\ref{MF}: \begin{eqnarray} \mathcal{L}^\prime \sim
\lambda^\prime \sum_l \psi^\dagger_l\psi_l (\vec{\phi})^2. \end{eqnarray}
When $\mu_1 = \mu_2 = 0$, this term is clearly irrelevant based on
straightforward power-counting. With finite Fermi surfaces, after
integrating out the fermions, this quadratic interaction will
induce the following four-body interaction between $\vec{\phi}$:
\begin{eqnarray} \mathcal{L}_2 = u
(\vec{\phi})^2_{i\omega,\vec{q}}\frac{|\omega|}{q}(\vec{\phi})^2_{-i\omega,-\vec{q}}
+ \cdots \label{l2}\end{eqnarray}
Note that the term with $|\omega|$ comes from the Fermi surface effects. 
We want to estimate the scaling dimension of $u$ at the 3d XY
fixed point. Since the scaling dimension $[(\vec{\phi})^2] = 3 -
1/\nu$, the scaling dimension of $u$ is $[u] = 2/\nu - 3$, here
$\nu$ is the standard critical exponent defined as $\xi \sim
r^{-\nu}$. Therefore as long as $\nu > 2/3$, $u$ is irrelevant.
This criterion is indeed satisfied according to the well-known
exponents of 3d O($N$) universality class \cite{vicari2003}.
$\mathcal{L}_2$ also exists at the $z = 2$ QCP with $\mu_1 \neq
\mu_2$ discussed before, however, since there $[\omega] = 2[q] =
2$, this term has a high scaling dimension and is clearly
irrelevant. A similar analysis about the $|\omega|/q$ term was
first made in Ref.~\cite{sachdevMorinari}.

Again we can evaluate effect of the coupling $\mathcal{L}^\prime$
on the Fermion self-energy. The leading order correction can again
be calculated through diagram Fig.~\ref{bosonself}$b$, while now
the dashed lines are correlation functions $ \langle
\vec{\phi}^2_{i\epsilon,\vec{k}} \vec{\phi}^2_{-i\epsilon,
-\vec{k}} \rangle \sim (\epsilon^2 + v_\phi^2 k^2)^{\frac{3}{2} -
\frac{1}{\nu}}$.  With finite Fermi surface, the leading order
contribution to the imaginary part of fermion self-energy reads
\begin{eqnarray} \Sigma(\omega)^{\prime\prime} \sim (\lambda^\prime)^2
|\omega|^{5 - 2/\nu}\mathrm{sgn}[\omega] \ll |\omega|. \end{eqnarray} Hence
this quadratic coupling $\mathcal{L}^\prime$ does not destroy the
Landau quasiparticle at the QCP either.

Now let us turn on an extra intra-edge Coulomb interaction in
Eq.~\ref{hamiltonian}: \begin{eqnarray} \mathcal{L}_v = \sum_{l}
Vn_{l,\uparrow}n_{l,\downarrow} \end{eqnarray} This term will favor to
develop magnetization on each edge \cite{yaoran2010}. Since
$Vn_{l,\uparrow}n_{l_\downarrow} \sim - V
(\psi^\dagger_{l}\sigma^z\psi_l)^2/2 $, the Hubbard-Stratonovich
transformation can give us mean field order parameter $\Phi_1 \sim
\langle \psi^\dagger \sigma^z \psi \rangle $ and $\Phi_2 \sim
\langle \psi^\dagger \sigma^z \eta^z \psi \rangle $ with Ising
symmetry. Without exciton condensate in the background, these two
Ising order parameters $\Phi_1$ and $\Phi_2$ are degenerate at the
mean field level. $\Phi_1$ breaks only $\mathcal{T}$, while
$\Phi_2$ breaks both $\mathcal{T}$ and $\mathcal{I}$. In the
background of exciton condensate, $\Phi_2$ has lower fermion mean
field energy because the exciton order parameters anticommute with
$\Phi_2$, hence the exciton condensate favors to have a
$\mathcal{I}$ breaking magnetization.

In Ref.~\cite{SMF}, it was shown that when $\mu_1 + \mu_2 = 0$, at
the vortex core of the exciton condensate order parameter there is
a Fermion zero mode, which carries charge $\frac{1}{2}$. This zero
mode is protected by the symmetry of Hamiltonian
Eq.~\ref{hamiltonian}: $ \gamma_2 H^\ast \gamma_2 = H $, and
$\gamma_2 = i\sigma^y\eta^y$. This symmetry guarantees that the
spectrum is symmetric with $E = 0$, and it is valid even with the
presence of exciton vortex. With nonzero $\Phi_2$, this symmetry
is broken, and there is no longer a zero mode at vortex core. By
contrast, if the system develops magnetization $\Phi_1$, there is
still a vortex core Fermion mode at precisely zero energy.

Just like the exciton order parameter $\phi$, the order parameter
$\Phi_a$ also couples to the Fermions both linearly and
quadratically. Due to the same matrix element suppression effect
as in Eq.~\ref{damp}, the linear coupling does not lead to
singular corrections to the QCP of $\Phi_a$. However, using the
similar argument as that below Eq.~\ref{l2}, the quadratic
coupling $ \mathcal{L}^\prime \sim \lambda \sum_l
\psi^\dagger_l\psi_l (\Phi_a)^2$ will lead to a relevant
perturbation at the 3d Ising fixed point as long as one of the
edges has a finite Fermi surface, due to the fact $\nu < 2/3$ at
the 3d Ising universality class \cite{vicari2003}.

In addition to the Fermi surface, the Goldstone mode of the
exciton condensate couples to order parameter $\Phi_a$ as well, if
the QCP of $\Phi_a$ occurs in a background of exciton condensate.
In the case without $\mathcal{I}$ ($\mu_1 \neq \mu_2$), the lowest
order coupling reads $\mathcal{L}^{\prime\prime} \sim
\lambda^{\prime\prime} (\partial_\tau\theta)(\Phi_a)^2$. $\theta$
is the phase angle of the exciton condensate: $\phi \sim
e^{i\theta}$. With $\mathcal{L}^{\prime\prime}$, after integrating
out the Goldstone mode $\theta$, a singular term is induced for
$\Phi_a$: \begin{eqnarray} \mathcal{L}_3 = u_3
(\Phi_a)^2_{i\omega,\vec{q}}\frac{\omega^2}{\omega^2 + v_\phi^2
q^2}(\Phi_a)^2_{-i\omega,-\vec{q}}. \label{l3}\end{eqnarray} To determine
the scaling dimension of this term, we again have to compare $\nu$
of 3d Ising transition and $2/3$: since $\nu < 2/3$, this coupling
$\mathcal{L}_3$ is also relevant at the 3d Ising universality
class. This relevant perturbation exists even when the Fermions
are fully gapped out by the exciton condensate. In the case with
$\mathcal{I}$, the coupling $\mathcal{L}^{\prime\prime}$ is
forbidden, since $\theta \rightarrow - \theta$ under
$\mathcal{I}$. In this case the coupling between $\Phi_a$ and the
exciton Goldstone mode will occur at higher order, hence no
relevant perturbation is induced at the 3d Ising universality
class.

Although the exciton is charge neutral, its transport
effect has been verified in bilayer quantum Hall
system~\cite{eisenstein}, by measuring the tunnelling conductance
between the two layers~\cite{MacDonald}. A similar measurement can
in principle be carried out in the thin film topological
insulator. The exciton condensate will lead to a sharp peak of the
inter-surface tunnelling conductance. Inside the exciton
condensate phase, the condensate will be destroyed by the thermal
fluctuation through a Kosterlitz-Thouless transition at finite
temperature. The scaling between the critical temperature $T_c$ of
this KT transition and the tuning parameter $r$ depends on the
universality class of the QCP, and $r$ can be taken as the
interaction $U - U_c$. For example, for the $z = 2$ mean field
transition, $T_c \sim |r|$; while for the 3D XY transition in the
phase diagram (Fig.~\ref{phasediagram}), $T_c \sim |r|^{z\nu} \sim
|r|^{2/3}$. Thus different quantum critical behaviors can be
measured through $T_c$. The interaction $U$ between the two
surfaces can be tuned by changing the thickness of the thin film
sample.

If $U$ in Eq.~\ref{hamiltonian} is attractive instead of
repulsive, then the system favors to have superconductor pairing.
After a particle-hole transformation for $\psi_2$: $\psi_2
\rightarrow \sigma^x \psi_2^\dagger$, both $U$ and $\mu_2$ change
sign, while all the other terms of the Hamiltonian remain
unchanged. The most energetically favored pairing state is
$\psi_1\sigma^x \psi_2$, because after particle-hole
transformation this pairing becomes the exciton condensate $\phi$.
Therefore all the analysis of the QCP and Goldstone mode about
this superconducting state can be obtained by particle-hole
transformation of the exciton case. For instance, at chemical
potential $\mu_1 = \mu_2$, there is a logarithmic divergence of
pairing susceptibility, while at $\mu_1 = - \mu_2$ there is a
matrix element suppression at the interaction vertex between
Cooper pair and fermions. The pairing $\psi_1\sigma^x \psi_2$ is a
spin triplet pairing with total $S^z = 0$, and the vortex core of
this superconductor carries a fermion zero mode when $\mu_1 =
\mu_2$.

In summary, we have studied the exciton condensation phase
transition and its quantum critical properties in a phase diagram
with edge dependent chemical potentials and Coulomb interaction.
Interplay between exciton condensate and other order parameters
are also discussed. In addition to the TBI materials that are
currently under intensive experimental studies, we expect our
formalism to be applicable to TBI with strong correlation, for
instance the materials with 5$d$ electrons
\cite{balents2009,nagaosa2008} which have been proposed recently.

\acknowledgments
The authors appreciate the very helpful discussions with Subir
Sachdev. Eun Gook Moon is sponsored by the National Science
Foundation under grant DMR-0757145, and also supported by the
Samsung Scholarship.

\end{document}